\documentclass[twocolumn,superscriptaddress,prl,amsmath,amssymb]{revtex4}
\usepackage{graphicx}
\let\xxxhat\hat
\renewcommand{\hat}[1]{{\mbox{\boldmath $ {\xxxhat {#1}} $}}}
\renewcommand{\vec}[1]{\mbox{\boldmath $ {#1} $}}

\begin{document}
\title{
Tomographic femtosecond X-ray diffractive imaging
}
\author{K. E. Schmidt}
\affiliation{Department of Physics, Arizona State University,
Tempe, AZ 85287-1504, USA}
\author{J. C. H. Spence}
\affiliation{Department of Physics, Arizona State University,
Tempe, AZ 85287-1504, USA}
\author{U. Weierstall}
\affiliation{Department of Physics, Arizona State University,
Tempe, AZ 85287-1504, USA}
\author{R. Kirian}
\affiliation{Department of Physics, Arizona State University,
Tempe, AZ 85287-1504, USA}
\author{X. Wang}
\affiliation{Department of Physics, Arizona State University,
Tempe, AZ 85287-1504, USA}
\author{D. Starodub}
\affiliation{Department of Physics, Arizona State University,
Tempe, AZ 85287-1504, USA}
\author{H. N. Chapman}
\affiliation{CFEL, DESY, Notkestra{\ss}e 85 22607 Hamburg, Germany}
\author{M. R. Howells}
\affiliation{ESRF, BP 220, 38043 Grenoble Cedex 9, France}
\author{R. B. Doak}
\affiliation{Department of Physics, Arizona State University,
Tempe, AZ 85287-1504, USA}

\begin{abstract}
A method is proposed for obtaining three simultaneous projections of a
target from a single radiation pulse, which also allows the relative
orientation of successive targets to be determined. The method
has application to femtosecond X-ray diffraction, and does not require
solution of the phase problem. We show that the principal axes of a
compact charge-density distribution can be obtained from projections
of its autocorrelation function, which is directly accessible in
diffraction experiments. The results may have more general application to
time resolved tomographic pump-probe experiments and time-series imaging.
\end{abstract}


\maketitle

Unless crystallographic redundancy can be taken advantage of, radiation
damage provides a well established limit to resolution for imaging in
biology.
X-ray microscopy of proteins is limited in this way to 10 nm\cite{howells2008}.
Since dose depends on the inverse fourth power of
resolution, a severe penalty attends any attempt to improve resolution
beyond this barrier, which occurs when the required dose needed to distinguish
adjacent image
voxels with statistical significance exceeds the damage limit at that
resolution (voxel size).

It has been suggested that the development
of the free-electron X-ray laser (FEL) may break this nexus between
dose and resolution \cite{neutze2000}, if it can provide sufficient photons for a
useful diffraction pattern in a single pulse, which terminates prior
to any of the characteristic times for damage
processes. A continuum of such times, femtosecond for electrons,
hundreds of femtoseconds for nuclear motion, is associated with
the various irreversible damage mechanisms and excitations. Theory
suggests that pulses shorter than the Auger decay time of a few
femtoseconds may terminate before significant disruption of the
valence electron distribution occurs. Experimental evidence, using
a 25 fs pulse of soft (30 nm) X-rays, now exists for this process of
``diffraction-before-destruction'' at low (90 nm) \cite{chapman2006b}
and higher periodically
averaged \cite{hauriege2007} resolution.
Since the FEL generates in excess of $10^{12}$
fully coherent photons in such a pulse, the method of diffractive
lensless imaging \cite{miao1999}, in which real-space images are reconstructed
computationally from these scattering patterns, would appear to provide
a means of overcoming the radiation damage barrier to high resolution
imaging in biology \cite{spence2007}.
The shortest FEL wavelength is currently 7 nm at
the Hamburg FLASH facility.  Shorter wavelength
FELs are being planned and constructed at other sites around the
world.

The subsequent destruction of the sample following the
initial elastic scattering event, however, has precluded
the possibility of
{\em three-dimensional} (tomographic) imaging of unique structures.
To overcome this limitation, we suggest here a means
to determine the relative
orientation of successive targets.

\begin{figure}
\includegraphics[width=7cm]{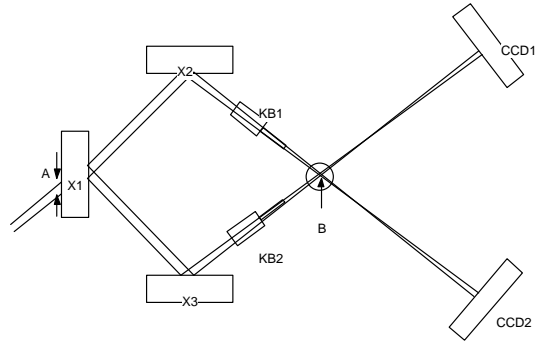}
\caption{
Scheme for tomographic femtosecond diffraction, drawn for only two beams
for simplicity.
Beamsplitter X1 is set to the dynamical 3-beam diffraction condition. 
Crystals X2 and X3 operate at the 2-beam dynamical condition. KB1 and KB2
are focusing optics for the target at B, with area
detectors CCD1 and CCD2.
}
\label{f1a}
\end{figure}
\begin{figure}
\includegraphics[width=7cm]{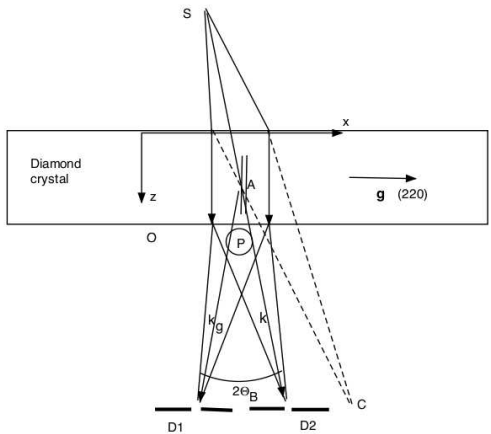}
\caption{
Two-beam beamsplitter with sample shown at P lying on the exit face of
the beamsplitter. The source S is focused onto two area detectors D1
and D2 containing central beam-dump holes. The two vertical arrows show
the direction of the Poynting vector. Three such orthogonal diffracted
beams, rather than the two shown, are proposed in the text.
}
\label{f1b}
\end{figure}
Two possible arrangements are shown in Figs. 1 and 2.
In Fig. \ref{f1a},
a beam-splitter and reflecting crystals
direct three orthogonal beams onto a non-periodic target particle
producing three far-field diffraction
patterns prior to destruction of the target.
All three two-dimensional
patterns are read out after each X-ray pulse, whereupon a new, identical
target such as a biomolecule is inserted in a new orientation.

We will show that the relative orientation of successive targets
can be determined
even if
the structure of the target is
unknown.

Following related work by  Smilgies on crystals \cite{smilgies2008}, we propose
experimental determination of the principal axes of the molecule to
describe its orientation relative to
the laboratory frame
defined by the incident probe beams. Multiple
scattering will be neglected, i.e., the first Born approximation
is assumed valid, so that the patterns have inversion symmetry and the target
density is a real function.

To expose the principle of the method, assume that
the phase problem can be solved, by, for example, iterative methods
(see \cite{marchesini2007}
for a review). (We will relax this assumption later.)
Then, in the ``flat'' Ewald sphere approximation,
i.e. at high probe energy, each beam delivers a projection in real space
(along the direction of the corresponding beam) of the scattering strength
per unit volume.
The projections will
be referred to different (randomly positioned) origins, and both
enantiomorphs (related by inversion symmetry) will be present with
equal likelihood. However once a particular enantiomorph is chosen for
one projection, the resulting two-dimensional envelope will constrain
the choice of enantiomorph for the other two projections.

Consider the
moments of the mass density $\rho(\vec r )$ for the target\cite{goldstein1950}.
The zeroth moment
delivers the total mass, the first moment delivers the center of mass
vector, and the second moment delivers the moment of inertia tensor. By
diagonalising the latter, the principal axes of the target may be found and
hence its orientation relative to the lab frame. Taking the center of
mass position as the origin, the inertia tensor is
\begin{equation}
\vec I = \iiint \rho(\vec r) \left (r^2\vec E-\vec r \vec r\right )
d\vec r \,,
\end{equation}
where $\vec E$ is the unit tensor and $\vec r \vec r$
is the outer product of the
position vector with itself. As with any symmetric tensor, $\vec I$
has only six independent elements, real eigenvalues, and orthogonal
eigenvectors corresponding to different eigenvalues.

Now take
$\rho(\vec r)$ to be the electronic charge
density of the target, whose projections
in three orthogonal directions are provided by the phased data, and which
define the $x$, $y$, and $z$ directions specified by unit vectors $\vec
e_i$ in the lab reference frame. The six independent elements of this
charge density
``inertia'' tensor then have the form
\begin{eqnarray}
\label{eq.2}
I_{zz} &=& \iint \rho_z(x,y) (x^2+y^2) ~dx dy
\nonumber\\
I_{xy} &=& -\iint \rho_z(x,y) xy ~dx dy
\end{eqnarray}
and similarly for $I_{xx}$, $I_{yy}$, $I_{yz}$ and $I_{xz}$.
Here $\rho_\alpha$ is 
the
projected density along the $\alpha$-direction.
Two of these six tensor elements
can be computed from each of the three projections, e.g. $I_{zz}$ and
$I_{xy}$
from the projection along the $z$-axis. Hence the inertia tensor of the
target is fully specified by computing moments and products of inertia
from the three projections. This charge density ``inertia''
tensor 
will differ from that based on mass
but serves equally well to
provide a consistent set of principal axes
fixed to the molecule and defining alignment. Being symmetric,
this tensor may be diagonalized by solving
the eigenvalue equations
\begin{equation}
\vec I \cdot \vec B = b \vec B
\end{equation}
for the three eigenvalues $b$ and corresponding eigenvectors $\vec B$.
These eigenvectors define a
new orthogonal coordinate system $\vec e'_j$
in which the three unit vectors lie
along the principal axes of the inertia tensor.
Barring degeneracy among the eigenvalues,
the three eigenvectors are unique to within a sign,
and therefore offer a natural means of specifying the orientation of the
target relative to the incident beam directions (lab frame). With the
unit vectors $\vec e_i$ of the lab frame and $\vec e'_j$
both known, the angles between
the principal axes of the target and the lab frame can immediately be
computed.
Thus the orientation of the target
has not only been defined by introducing the principal axes of the
inertia tensor, but also specified (within polarity) relative to the lab
coordinates.

Clearly, then,
to establish the orientation
of the particle it suffices to
(i) Record three diffraction patterns,
one for each of the three incident beam directions.
(ii) Invert the
diffraction patterns using phase retrieval techniques to yield three
real-space projections of the scattering strength.
(iii) Compute the first moment of each projection to obtain the center of
mass position for that projection.
(iv) Compute the second order moments of
each projection (products of inertia) about the center of mass to obtain
one diagonal and one off-diagonal tensor element.
(v) Diagonalize the
resulting tensor to obtain the eigenvectors of the tensor.
(vi) Compute
the orientation of each beam relative to the eigenvectors of the target
in order to determine the angles between laboratory and principal
axes coordinates.
(vii) If this process is repeated for many successive
identical targets in random orientations, their relative orientations
can be found, and hence a complete three-dimensional tomographic image
can be assembled by standard tomographic techniques such as filtered
backprojection.

We now extend this analysis to show
that the principal axes may be found, {\em even without}
solving the phase problem, by working with the autocorrelation of the
sample density
\begin{eqnarray}
A(\vec r) = \int d\vec r' \rho(\vec r+ \vec r')\rho(\vec r') \,.
\end{eqnarray}
A typical product of inertia is
\begin{eqnarray}
I^A_{xy} = -\int d\vec r ~x y A(\vec r) =
-\int d\vec r'\rho(\vec r')
\left [ \int d\vec r x y \rho(\vec r+ \vec r') \right ]
\nonumber\\
= \int d\vec r' \rho(\vec r') \left [ I^\rho_{xy}-x'y' M\right ]
= 2MI^\rho_{xy}
\end{eqnarray}
where $M=\int d\vec r\rho(\vec r)$, and
with use of
the parallel axis theorem to calculate the product of
inertia for the shifted coordinates. Therefore,
the principal axes of the autocorrelation
function are the same as the principal axes of the corresponding density.

Given a ``flat'' Ewald sphere,
the Fourier
transform of each diffraction pattern (intensity) directly provides a projection
of the three-dimensional autocorrelation function of the density, and
the analysis simply requires changing $\rho(\vec r)$
to $A(\vec r)$ in Eqs. \ref{eq.2}.

More generally,
the moment of inertia can be calculated from the second derivative of
the Fourier transform. Denoting these by a tilde,
\begin{eqnarray}
\tilde \rho(\vec q) &=& \int d\vec r e^{-i \vec q \cdot \vec r}\rho(\vec r)\,,
\ \tilde A(\vec q) = \tilde \rho(\vec q) \tilde \rho(-\vec q)
\nonumber\\
I^A_{xy} &=& \left . \partial_{q_x}\partial_{q_y} \tilde A(\vec q)
\right |_{\vec q = 0}
=
\left . \partial_{q_x}\partial_{q_y} \tilde \rho(\vec q) \tilde \rho(-\vec q)
\right |_{\vec q = 0}
\nonumber\\
&=& \left .\tilde\rho(0)[2\partial_{q_x}\partial_{q_y} \tilde\rho(\vec q) ]
\right |_{\vec q= 0} = 2MI^\rho_{xy}
\end{eqnarray}
as before. Replacing $\tilde A(q_x,q_y,0)$, (``flat'' Ewald sphere)
with the correct Ewald sphere (finite radius) diffraction pattern,
and taking the incident wave vector $\vec k$ along $z$,
\begin{eqnarray}
\left . \partial_{q_x}\partial_{q_y}\tilde A
\left (q_x,q_y,\sqrt{k^2-q_x^2-q_y^2}-k\right ) 
\right |_{q_x=q_y=0} =
\nonumber\\
\left . \partial_{q_x}\partial_{q_y}\tilde A(q_x,q_y,0) 
\right |_{q_x=q_y=0}
\end{eqnarray}
so that the moments of the Fourier transform of the diffracted data
still yield the same principal axes.

\begin{figure}
\includegraphics[width=8cm]{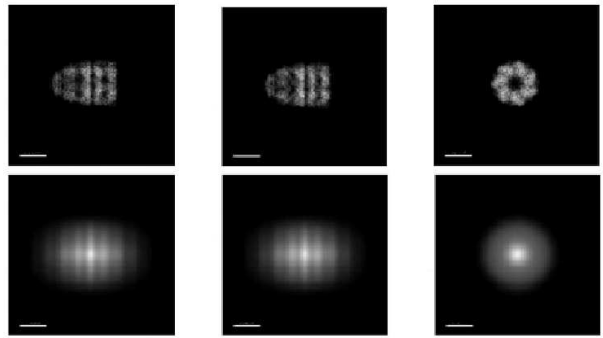}
\caption{
The three orthogonal projections of the GroEL charge
density (upper) and the corresponding projections of the autocorrelation
function (lower). The bar indicates 10 nm.}
\label{f2}
\end{figure}
\begin{figure}
\includegraphics[width=8cm]{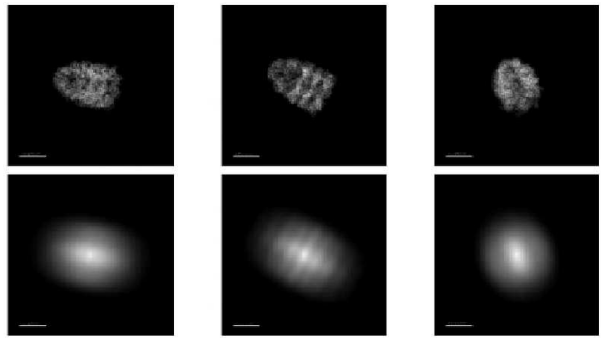}
\caption{
Projections of GroEL density (upper) and autocorrelation function (lower)
in a second random orientation.
}
\label{f3}
\end{figure}
We have investigated this procedure using detailed
numerical simulations based on data in the Protein Data base for
GroEL-GroES protein complex (PDB entry 1SVT).  The
three-dimensional density was synthesised from the tabulated atomic
coordinates. Fig. \ref{f2} shows the projected densities and corresponding
projected autocorrelation functions using the principal axes obtained
from Eqs. \ref{eq.2}. A second density was then generated in a random
orientation with respect to the first, as shown in Fig. \ref{f3}. For
each of these orientations the principal axes were determined using
both the densities and the autocorrelation functions, giving similar
results. 
The rotation matrix needed to rotate from the first (Fig. \ref{f2}) to 
the second (Fig. \ref{f3}) orientation was generated from the principal 
axes. As a result of inversion symmetry there are four distinct 
choices of rotation matrix (corresponding to choices of 
eigenvector signs) when the autocorrelation function is used.
The correct rotation matrix was obtained by testing each to see
which predicted lines of intensity in diffraction patterns
common to two orientations.
(Any two planes in reciprocal space passing through
the origin must intersect along a common line). In this way only one
rotation matrix will be found to give consistent results. Numerical
trials have found this procedure to be reliable with several different
test objects.
Our use of specific common lines should be more robust for noisy data
than common line search methods.  (Shneerson et al. \cite{shneerson2008}
have shown that identifications of common lines in diffraction pattern
down to a mean photon count of 10 per pixel enables the determination of
their relative orientations without the need to solve the phase problem.)

This treatment easily extends to
the case where the beams
from the beamsplitter are not orthogonal. Reciprocal vectors can be
defined in the usual way, so that each pattern lies in the plane of
two of these vectors. The products of inertia may be simply evaluated
in terms of these reciprocal vectors, and finally transformed into the
required lab frame moments.

Two separate experimental
implementations of this approach are suggested in Figs. \ref{f1a} and
\ref{f1b},
each with an incident beam close to the [111] direction of a diamond
beam-splitter crystal, set to simultaneously excite the [022] and
[2$\overline{2}$0] reflections. For an X-ray energy close to 7 keV, this generates
three orthogonal beams with Bragg angles of $45^{\circ}$. The Borrmann
effect \cite{batterman1964} may then be used to produce three beams of approximately equal intensity for a crystal thickness of order one mm, due to 3-beam multiple
scattering \cite{umeno1975,umeno1975x,zuo1989}.
This remarkable effect, in which wavefield
components with zero-crossings at atom positions avoid photoelectron
production, reduces absorption in the beamsplitter by many orders of
magnitude. Crystals X1 and X2 operate at 2-beam dynamical conditions, for
which reflectivities of greater than 90\% are possible.  For the arrangement
in Fig. \ref{f1a}, the important experimental challenge is to generate
three incident beams that converge focused
onto a micron-sized volume of space at
the same instant in time. ``Diffraction before destruction'' will require
pulse durations of 10 fs or less, corresponding to a spatial pulse length
of at most 3 $\mu$m, a technically challenging but feasible length
scale for experimental realization. (Calculations show that
8 keV x-ray pulses reflected from Si(111) are stretched
by about 4 fs\cite{shastri2001}.)

A monolithic integration of this arrangement
may be possible. The arrangement in Fig. \ref{f1b}, with sample mounted on
the beamsplitter, provides isochronal, but unfocused,
optical paths
to the sample.
This arrangement is better suited
to long exposures of continuous radiation for stationary samples. We do
not provide detailed phase-space matching calculations here, but note
that the beam divergence of one planned FEL, the Linac
Coherent Light Source at Stanford, USA, is $1.1 \times 10^{-6}$
radians, which is less
than a typical perfect-crystal rocking curve width of
$3.4 \times 10^{-5}$ radians. An
energy spread of $\Delta E/E = 1.4 \times 10^{-4}$
can be expected after monochromation
at 8 kV, with a beam width of 20 $\mu$m. Estimates suggest
that even with less than one scattered photon per pixel,
phasing and reconstruction is
possible\cite{shneerson2008,elser2007,ourmazd2008}.

We conclude that a determination of the relative orientation between
successive particles of unknown structure (each initially in a random
unknown orientation with respect to the laboratory frame) may be
achieved without the need to solve the phase problem.

For a
stream of identical molecules in random orientations, this
would allow data from different molecules to be merged in the correct
relative orientation. After phasing the resulting three-dimensional
reciprocal-space data, a tomographic image can then be reconstructed. We
have also shown that the orientation of successive objects can be
determined from autocorrelation functions, so that a solution of
the phase problem is not required.
The
entire procedure cannot distinguish enantiomorphs. Stereoscopic
projections might be obtained from just two projections.
We have suggested experimental
implementations for this method for femtosecond X-ray diffraction.
This analysis
applies to any penetrating particles (e.g., neutrons or high energy
electrons), insofar as the scattering can be characterized by a scalar
potential and the orientation and structure of the sample is unknown
(unlike goinometer-based systems where both coordinate systems are
known).
This might include, for example, the tracking of the orientation
of a single body from which nondestructive diffraction patterns can
be obtained as function of time.
``Proof of principle'' measurements at optical wavelengths are currently
under way.

\begin{acknowledgments}
This work was supported by NSF IDBR 0555845
and CBST at UC Davis.
\end{acknowledgments}
\vskip -2mm

\bibliographystyle{apsrev}
\bibliography{beam}

\end{document}